\documentclass[twocolumn,amsmath,amssymb,floatfix,prb,shownopacs,footinbib,superscriptaddress]{revtex4}
\usepackage{amsmath,amssymb,natbib,bm,graphicx,url}
\usepackage[british]{babel}

\begin{document}

\title{Non-Equilibrium Edge Channel Spectroscopy in the Integer Quantum Hall Regime}
\author{C. Altimiras}
\affiliation{CNRS, Laboratoire de Photonique et de Nanostructures
(LPN) - Phynano team, route de Nozay, 91460 Marcoussis, France}
\author{H. le Sueur}
\affiliation{CNRS, Laboratoire de Photonique et de Nanostructures
(LPN) - Phynano team, route de Nozay, 91460 Marcoussis, France}
\author{U. Gennser}
\affiliation{CNRS, Laboratoire de Photonique et de Nanostructures
(LPN) - Phynano team, route de Nozay, 91460 Marcoussis, France}
\author{A. Cavanna}
\affiliation{CNRS, Laboratoire de Photonique et de Nanostructures
(LPN) - Phynano team, route de Nozay, 91460 Marcoussis, France}
\author{D. Mailly}
\affiliation{CNRS, Laboratoire de Photonique et de Nanostructures
(LPN) - Phynano team, route de Nozay, 91460 Marcoussis, France}
\author{F. Pierre}
\email[email: ]{frederic.pierre@lpn.cnrs.fr}
\affiliation{CNRS, Laboratoire de Photonique et de Nanostructures
(LPN) - Phynano team, route de Nozay, 91460 Marcoussis, France}
\date{October 14, 2009}

\maketitle

{\sffamily Heat transport has large potentialities to unveil new physics in mesoscopic systems. A striking illustration is the integer quantum Hall regime \cite{vklitzing1980iqhe}, where the robustness of Hall currents limits information accessible from charge transport \cite{fertig2009vfe}. Consequently, the gapless edge excitations are incompletely understood. The effective edge states theory describes them as prototypal one-dimensional chiral fermions \cite{halperin1982qhc,wen1992tesfqh} - a simple picture that explains a large body of observations \cite{buttiker1988abs} and calls for quantum information experiments with quantum point contacts in the role of beam splitters \cite{Ionicioiu2001qcwbe,stace2004mowc,ji2003emz,feve2007ses,samuelsson2004twoeAB,olkhovskaya2008hom}. However, it is in ostensible disagreement with the prevailing theoretical framework that predicts, in most situations \cite{chklovskii1992eec}, additional gapless edge modes \cite{aleiner1994nee}. Here, we present a setup which gives access to the energy distribution, and consequently to the energy current, in an edge channel brought out-of-equilibrium. This provides a stringent test of whether the additional states capture part of the injected energy. Our results show it is not the case and thereby demonstrate regarding energy transport, the quantum optics analogy of quantum point contacts and beam splitters. Beyond the quantum Hall regime, this novel spectroscopy technique opens a new window for heat transport and out-of-equilibrium experiments.}

The integer quantum Hall effect, discovered nearly thirty years ago \cite{vklitzing1980iqhe}, has recently experienced a strong revival driven by milestone experiments toward quantum information with edge states \cite{ji2003emz,feve2007ses,neder2007i2e}. Beyond Hall currents, new phenomena have emerged that were unexpected within the free one-dimensional chiral fermions (1DCF) model. The on-going debate triggered by electronic Mach-Zehnder interferometers experiments \cite{ji2003emz,roulleau2008lphi,litvin2007dsec,bieri2009fbv} vividly illustrates the gaps in our understanding. Coulomb interaction is seen as the key ingredient. In addition to its most striking repercussion, the fractional quantum Hall effect \cite{tsui1982fqhe}, the edge reconstruction (ER) turns out to have deep implications on edge excitations. This phenomenon results from the competition between Coulomb interaction that tends to spread the electronic fluid, and the confinement potential: as the latter gets smoother, the non-interacting edge becomes unstable \cite{macdonald1993er}. Theory predicts new branches of gapless electronic excitations in reconstructed edges \cite{aleiner1994nee,chamon1994er}, which breaks the mapping of an edge channel (EC) onto 1DCF and, possibly, the promising quantum optics analogy. For most edges realized in semi-conductor heterojunctions (except by cleaved edge overgrowth \cite{chang1996ocl}), ER results in wide compressible ECs separated by narrow incompressible strips \cite{chklovskii1992eec} and the new excited states are overall neutral internal charge oscillations across the ECs width \cite{aleiner1994nee}.
\begin{figure}[bt]
\includegraphics[width=1\columnwidth,clip]{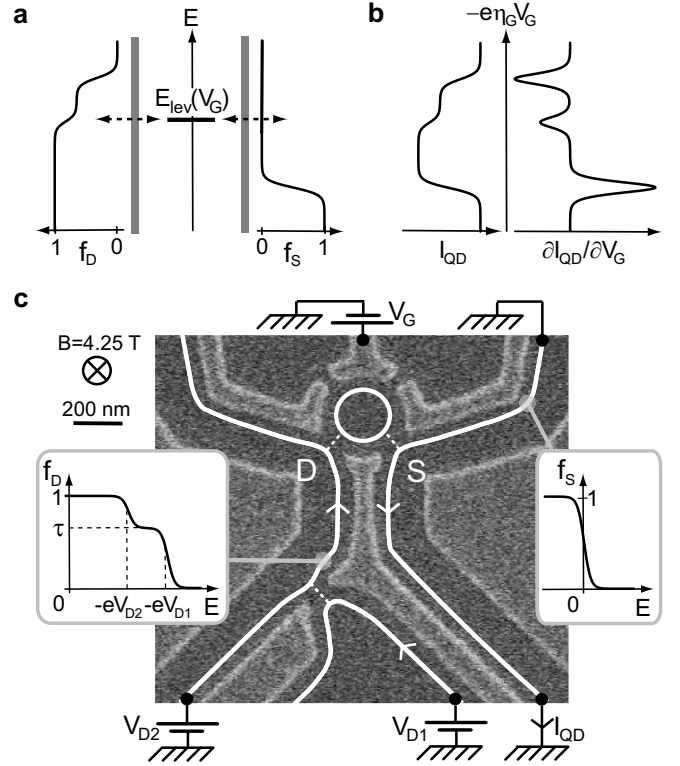}
\caption{{\bfseries Experimental implementation of non-equilibrium edge channel spectroscopy. a}, Schematic description of the energy distributions $f_{D,S}(E)$ spectroscopy with a single active electronic level of tunable energy $E_{lev}(V_G)$ in the quantum dot (QD). {\bfseries b}, The current $I_{\mathrm{QD}}$ ($\partial I_{\mathrm{QD}}/\partial V_G$) is proportional to $f_S(E)-f_D(E)$ ($\partial (f_S(E)-f_D(E))/\partial E$) ignoring variations in tunnel rates and tunneling density of states. {\bfseries c}, E-beam micrograph of the sample. Surface metal gates appear brighter. Electronic excitations propagate counter clockwise along two edge channels (EC) of the quantum Hall regime. The outer EC (continuous white lines) is partly transmitted (dashed lines) across quantum point contact (QPC) and QD. The inner EC (not shown) is always reflected. The QPC is used to drive out-of-equilibrium the drain outer EC. Gates partly covered by the insets are grounded and do not influence the electron paths. Left inset: non-interacting electrons prediction for $f_D(E)$ in the outer EC at output of QPC.} \label{Fig1}
\end{figure}

In practice, the predicted additional neutral modes are transparent to Hall currents. More surprisingly, a linear I-V characteristic is frequently observed for tunnel contacts (different behaviors were also reported, e.g. \cite{roddaro2005phs}) whereas a non-linear characteristic is predicted \cite{conti1996cme,han1997dce,yang2003ftd}. This contradiction is resolved by assuming \textit{ad-hoc} that only \textit{rigid displacements} of compressible ECs are excited by tunnel events, and not internal excitations \cite{aleiner1994nee,conti1996cme,han1997dce,zulicke1999pdt}. The rigid displacement model arguably relies on the overriding strength of Coulomb interaction that tends to orthogonalize bare tunneling electrons and correlated electronic fluids \cite{zulicke1999pdt}. However, the above argument does not hold at arbitrary transmission probabilities, where multiple electrons processes occur. Therefore, the role of predicted internal excitations has to be determined experimentally. The present work provides such a test. An EC is driven out-of-equilibrium with a quantum point contact (QPC) of arbitrary transmission, possibly exciting internal modes. A short distance away, the resulting energy distribution $f(E)$ is measured with a tunnel coupled quantum dot (QD) expected to probe only rigid displacement excitations, hereafter called quasiparticles. Consequently, the amount of energy injected into internal modes at the QPC would appear as an energy loss in $f(E)$.

Measurements of the energy distribution in mesoscopic devices were first performed in 1997 on metallic circuits using a superconducting tunnel probe \cite{pothier1997relax}. More recently, the regime of high magnetic fields was accessed using dynamical Coulomb blockade \cite{anthore2003relaxdcb}. In two-dimensional electron gas systems, non-Fermi energy distributions could not be measured since transferring the techniques developed for metal circuits is technically challenging (although hot electrons have been detected, e.g. \cite{heiblum1985dobt}). Regarding the quantum Hall regime (QHR), state of the art is the very recent qualitative probe of heating \cite{granger2008och}. Here, we demonstrate that $f(E)$ can be fully extracted from the tunnel current across a QD. In the sequential tunneling regime, the discrete electronic levels in a QD behave as energy filters \cite{averin1991tsec,kouwenhoven1997etqd}, as previously demonstrated with double QDs \cite{vandervaart1995rt}. Assuming a single active QD level of energy $E_{lev}$, and ignoring the energy dependence of tunnel rates and tunneling density of states in the electrodes, the QD current reads
\begin{eqnarray}
I_{\mathrm{QD}}=I_{\mathrm{QD}}^{max}(f_S(E_{lev})-f_D(E_{lev})), \label{Idot}
\end{eqnarray}
where the subscript $S$ ($D$) refers to the source (drain) electrode; $f_{S,D}$ are the corresponding energy distributions and $I_{\mathrm{QD}}^{max}$ is the maximum QD current. In practice, $f_{S,D}$ are obtained separately by applying a large enough source-drain voltage (Fig.~\ref{Fig1}a,b) and the probed energy $E_{lev}=E_0-e\eta_G V_G$ is swept using a capacitively coupled gate biased at $V_G$, with $\eta_G$ the gate voltage-to-energy lever arm and $E_0$ an offset. Raw data $\partial I_{\mathrm{QD}}/\partial V_G$ measured by lock-in techniques are proportional to $\partial f_{D,S} (E)/\partial E$.

\begin{figure}[tb]
\includegraphics[width=1\columnwidth,clip]{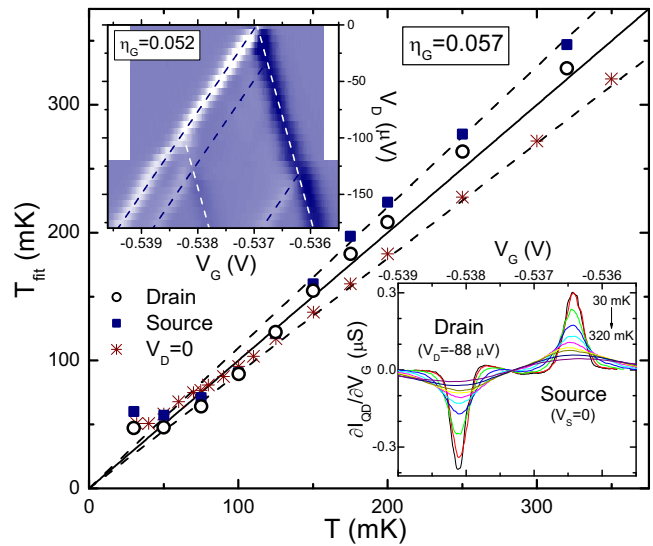}
\caption{{\bfseries Equilibrium edge channel spectroscopy and quantum dot characterization.} Temperatures $T_{fit}$ are obtained by fitting the QD-drain dip ($\circ$) and QD-source peak ($\blacksquare$) in $\partial I_{\mathrm{QD}}/\partial V_G$ (resp. left and right peaks in bottom right inset), and from the single $\partial I_{\mathrm{QD}}/\partial V_D$ peak at $V_D=0$ ($\ast$) (raw data not shown) with $\eta_G=0.057$. We assumed a single active level in the QD and Fermi energy distributions in source and drain. Errors within $T\pm10\%$ are enclosed between the dashed lines. Top left inset: Surface plot of $\partial I_{\mathrm{QD}}/\partial V_G$ (negative is brighter, positive darker) measured at $T=30$~mK for an outer (inner) drain EC biased at $V_D$ ($-88~\mu$V). The intense stripes slopes yield $\eta_G=0.052\pm9\%$. Enclosed dashed lines outline the small contributions of other electronic levels.} \label{Fig2}
\end{figure}

A tunable non-Fermi energy distribution is generated in an EC with a voltage biased QPC. Similar setups were used previously to create imbalanced electron populations between co-propagating ECs \cite{vanwees1989aiqhe}, each characterized by a cold Fermi distribution. Only in a very recent experiment \cite{granger2008och}, was an EC heated up. Beyond heating, $f(E)$ is here controllably tuned out of equilibrium. Let us consider one EC and assume it can be mapped onto non-interacting 1DCF. According to the scattering approach \cite{gramespacher1999noneqf}, the energy distribution at the output of a QPC of transmission $\tau$ is a tunable double step (Fig.~\ref{Fig1}c, left inset)
\begin{equation}
f_{D}(E)=\tau f_{D1}(E)+(1-\tau)f_{D2}(E),
\label{Eq2}
\end{equation}
where $f_{D1}$ $(f_{D2})$ is the equilibrium Fermi distribution function in the partially transmitted (reflected) incoming EC of electrochemical potential shifted by $eV_{D1}$ ($eV_{D2}$). In presence of ER, the above energy distribution applies to the quasiparticles if internal modes are not excited at the QPC. Else, there are no theoretical predictions because a QPC is very difficult to treat non-perturbatively in their natural bosonic formalism.

The sample shown in Fig.~\ref{Fig1}c was tailored in a two-dimensional electron gas realized in a GaAs/Ga(Al)As heterojunction, set to filling factor 2 and measured in a dilution refrigerator of base temperature 30~mK. The experiment detailed here focuses on the outer EC represented as a white line. The inner EC (not shown) is fully reflected by the QPC and the QD. We checked that charge tunneling between ECs is negligible along the 0.8~$\mu$m propagation length from QPC to QD.

\begin{figure}[tb]
\includegraphics[width=1\columnwidth,clip]{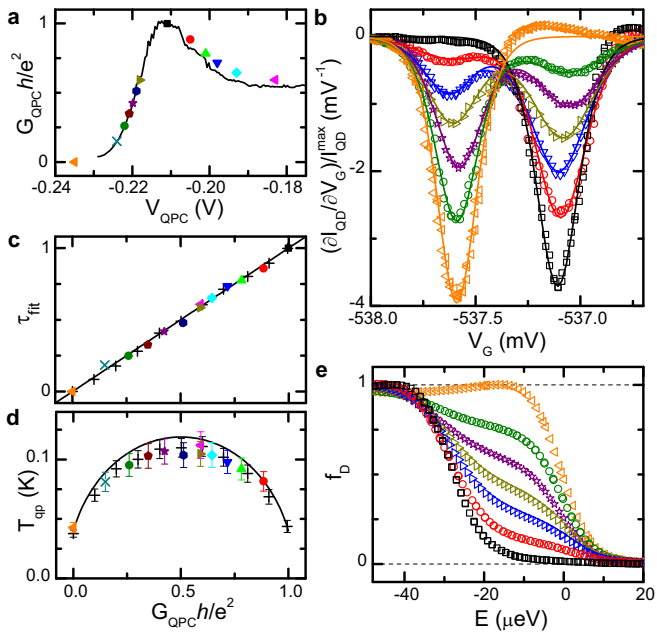}
\caption{{\bfseries Spectroscopy of an edge channel tuned out-of-equilibrium with the conductance of a QPC.} The QPC's voltage bias is here fixed to $\delta V_D \equiv V_{D1}-V_{D2}=36~\mu$V. Data shown as $(+)$ in ({\bfseries c}) and ({\bfseries d}) were obtained in a second cooldown with a renewed $\eta_G=0.059\pm7\%$. {\bfseries a}, Measured $G_{\mathrm{QPC}}$ versus $V_{\mathrm{QPC}}$ applied to the lower left metal gate in Fig.~\ref{Fig1}c. Symbols are data points obtained during the corresponding EC spectroscopy. The continuous line was measured earlier with $\delta V_D=0$. {\bfseries b}, Measured $\partial I_{\mathrm{QD}}/\partial V_G$ (symbols) for the QD-drain contribution (data have been aligned in $V_G$ and normalized to $I_{\mathrm{QD}}^{max}$). The two dips correspond to a double step energy distribution $f_D(E)$. {\bfseries c}, Symbols are $\tau_{fit}$ obtained from fits of the data (continuous lines in ({\bfseries b})) assuming $f_D(E)$ is the weighted sum of two Fermi functions. We find an accurate agreement with the non-interacting 1D chiral fermions (1DCF) model prediction $\tau_{fit}=G_{\mathrm{QPC}}h/e^2$. {\bfseries d}, Generalized non-equilibrium temperature (symbols, see text) extracted from the data and theoretical prediction for free 1DCF (continuous line). Error bars are dominated by uncertainties on $\eta_G$. {\bfseries e}, Energy distributions obtained by integrating the data in ({\bfseries b}).}
\label{Fig3}
\end{figure}

We first perform a standard non-linear QD characterization \cite{kouwenhoven1997etqd} (Fig.~\ref{Fig2}, top left inset). The two large signal stripes are frontiers of consecutive Coulomb diamonds and are accounted for by a single active QD level. Small contributions of three additional levels of relative energies \{-95,30,130\}~$\mu$eV are also visible. The lever arm extracted from the stripes' slopes is $\eta_G\simeq0.052\pm9\%$.

Then, we test the spectroscopy with known Fermi functions by measuring $\partial I_{\mathrm{QD}}/\partial V_G(V_G)$ at $V_{D1}=V_{D2}=-88~\mu V$ for several temperatures (Fig.~\ref{Fig2}, bottom right inset). By fitting these data with Eq.~\ref{Idot} using Fermi functions, we extract a fit temperature scaled by the lever arm $T_{fit}/\eta_G$. The value $\eta_G=0.057$, compatible with the non-linear QD characterization, is found to reproduce best the mixing chamber temperature $T$ with $T_{fit}$. The drain ($\circ$) and source ($\blacksquare$) fit temperatures are shown in Fig.~\ref{Fig2} together with $T_{fit}$ obtained using the standard procedure \cite{kouwenhoven1997etqd} from $\partial I_{\mathrm{QD}}/\partial V_D(V_G)$ at $V_D\simeq0$ ($\ast$). We find deviations mostly within $\pm10\%$ (dashed lines in Fig.~\ref{Fig2}) except for a saturation at $T_{fit}\approx50~$mK possibly due to a higher electronic temperature. In the following, we use $\eta_G=0.057$ obtained here in the same experimental configuration as to measure unknown $f(E)$s.

Electrons are now driven out-of-equilibrium in the drain outer EC. In the following, the electrode $D2$ and the inner drain EC are voltage biased at $V_{D2}=-88~\mu$V and the source ECs are emitted by a cold ground.

First, the bias voltage across the QPC is set to $\delta V_D \equiv V_{D1}-V_{D2}=36~\mu$V and its conductance $G_{\mathrm{QPC}}=\tau e^2/h$ is tuned by applying $V_{\mathrm{QPC}}$ to the bottom left gate in Fig.~\ref{Fig1}c (see Fig.~\ref{Fig3}a). Note that at 30~mK, we find the transmission $\tau$ is constant within $2\%$ with the QPC voltage bias below $36~\mu$V. Typical sweeps $\partial I_{\mathrm{QD}} /\partial V_G(V_G)$ and the corresponding $f_D(E)$ are shown in Figs.~\ref{Fig3}b and \ref{Fig3}e, respectively. The QD-drain negative contribution transforms from a single dip at $\tau=\{0,1\}$ into two dips separated by a fixed gate voltage and whose relative weights evolve monotonously with $\tau \in ]0,1[$. Continuous lines are fits with Eq.~\ref{Eq2} using for $f_{D1,D2}$ two Fermi functions shifted by a fixed energy and weighted by the factors $\tau_{fit}$ and $1-\tau_{fit}$. The values of $\tau_{fit}$ are found to deviate by less than 0.03 from the measured transmission $\tau$ (Fig.~\ref{Fig3}c), in accurate agreement with the free 1DCF model. Symbols ($+$) in Fig.~\ref{Fig3} correspond to data obtained in a second cooldown.

\begin{figure}[tb]
\includegraphics[width=1\columnwidth,clip]{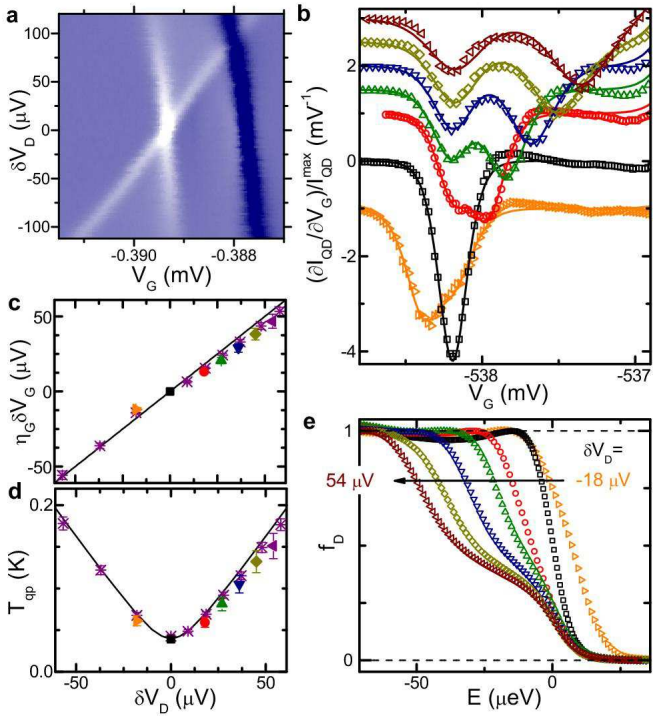}
\caption{{\bfseries Spectroscopy of an edge channel tuned out-of-equilibrium with the voltage across a QPC.} The QPC's conductance is here fixed to $G_{\mathrm{QPC}} \approx 0.5 e^2/h$. Data of ({\bfseries a}) and ($\ast$) in ({\bfseries c}) and ({\bfseries d}) were obtained in a third cooldown with a renewed $\eta_G=0.062\pm5\%$. {\bfseries a}, Surface plot of $\partial I_{\mathrm{QD}}/\partial V_G (\delta V_D,V_G)$. The QD-drain contribution (brighter) splits into two similar dips separated by a gate voltage difference proportional to $\delta V_D$. {\bfseries b}, Measured $\partial I_{\mathrm{QD}}/\partial V_{G}$ (symbols) for the QD-drain contribution. Data have been shifted vertically for clarity, and horizontally to align the peak corresponding to the fixed potential $V_{D2}$. Continuous lines are fits assuming $f_D(E)$ is the weighted sum of two Fermi functions. {\bfseries c}, Symbols are the fit parameters $\eta_G\delta V_{G}$. The continuous line is the prediction for non-interacting 1DCF. {\bfseries d}, Generalized non-equilibrium temperature (symbols) and theoretical prediction for free 1DCF (continuous line). The good agreement data-prediction demonstrates that internal modes are not excited at the QPC within our experimental accuracy. {\bfseries e}, Energy distributions obtained by integrating the data in ({\bfseries b}).}
\label{Fig4}
\end{figure}

In a second step, the QPC transmission is fixed to $\tau \approx 0.5$ and the bias voltage $\delta V_D$ is changed. Typical raw data are shown in Fig.~\ref{Fig4}a. These were obtained in a third cooldown with a QD renewed by the thermal cycle showing no signs of additional QD levels in the probed energy range. The single dip in the QD-drain contribution (bright) at $\delta V_D=0$ splits in two similar dips that are separated by a gate voltage difference proportional to $\delta V_D$. Meanwhile, the QD-source peak (dark) is mostly unchanged but slowly drifts parallel to one QD-drain dip due to the capacitive coupling between drain and QD. In the first cooldown, $V_{D1}$ was kept within $[-106,-34]~\mu$V to minimize complications related to additional QD levels (lower bound) and to ensure well-separated source and drain contributions (upper bound). Symbols in Figs.~\ref{Fig4}b and \ref{Fig4}e are, respectively, data and extracted $f_D(E)$ for the QD-drain contribution at $\delta V_D=\{-18,0,18,27,36,45,54\}~\mu$V and $\tau=0.58$. Continuous lines in Fig.~\ref{Fig4}b are fits with Eq.~\ref{Eq2} using the measured $\tau$ and for $f_{D1,D2}$ two Fermi functions shifted in energy by the fit parameter $-e\eta_G\delta V_{G}$. The resulting $\eta_G \delta V_{G}$ are plotted as symbols versus $\delta V_D$ in Fig.~\ref{Fig4}c. Those obtained in the third cooldown are shown as ($\ast$) using the renewed lever arm $\eta_G=0.062$. We find $\eta_G \delta V_{G} \simeq \delta V_D$ as expected from the non-interacting 1DCF model. Deviations are always smaller than $8~\mu$V ($5~\mu$V) for the first (third) cooldown, a reasonable agreement regarding uncertainties in $\eta_G$ of $\pm10\%$ ($\pm5\%$).

In the two experiments above, we found the measured quasiparticle $f(E)$s verify predictions of the scattering approach. In order to establish the analogy QPC-beam splitter one also needs to demonstrate that internal EC modes are not excited. A direct test consists in extracting the quasiparticle heat current $J_E^{qp}$ from the data, and comparing it with the full edge excitations heat current $J_E$ obtained from power balance considerations (see Supplementary Information for details):
\begin{equation}
J_E(T=0)=\frac{(e \delta V_D)^2}{2h}\tau (1-\tau).
\label{Eq3}
\end{equation}
The cancellation $v \nu=1/h$ of velocity ($v$) and density of states per unit length and energy ($\nu$) that applies to the 1DCF quasiparticles permits us to obtain $J_E^{qp}$ from the measured $f(E)$ without any sample specific parameters:
\begin{equation}
J_E^{qp}=\frac{1}{h}\int (E-\mu) (f(E)-\theta(\mu-E))dE,
\label{Eq4}
\end{equation}
with $\mu$ the electrochemical potential and $\theta(E)$ the step function. Consequently, \textit{we measure quantitatively the quasiparticle heat current}. The result of this procedure is shown as symbols in Figs.~\ref{Fig3}d and \ref{Fig4}d using the generalized non-equilibrium temperature $T_{qp} \equiv \sqrt{6 h J_E^{qp}}/\pi k_B$ together with the prediction $T_{1DCF}=\sqrt{T^2+\tau(1-\tau)3(e \delta V_D/\pi k_B)^2}$ if none of the injected power is carried on by internal modes (continuous lines). We find a good agreement $T_{qp}\simeq T_{1DCF}$ without fitting parameters and essentially in or close to error bars. Hence, within our experimental accuracy, the propagative internal modes do not contribute to heat transport and therefore are not excited. Note that the relatively small observed deviations are cooldown dependent, which points out the QD. Indeed, the data can be more accurately accounted for including a second active QD level (see Supplementary Information). Last, preliminary data show a significant energy redistribution with the inner EC for propagations longer than $2~\mu$m in the probed energy range. Therefore, the observed small discrepancies could also result from the finite $0.8~\mu$m propagation length.

Overall, we demonstrate that QPCs in the QHR are tunable electrical beam splitters for one-dimensional fermions, i.e. rigid edge channel displacements, \textit{i)} by comparing the energy distribution at a QPC output with predictions of the scattering approach \cite{buttiker1988abs}, and \textit{ii)} by showing that internal edge channel modes are not excited. This does not only rule out non-ideal QPC behaviors to explain the surprising phenomena observed on electronic Mach-Zehnder \cite{ji2003emz,roulleau2008lphi,litvin2007dsec,bieri2009fbv}. It also establishes a solid ground for future quantum information applications with edge states. Finally, an essential part of this work is the demonstration of a new technique to measure the fundamental energy distribution function. It makes $f(E)$ accessible for most systems were quantum dots can be realized. We expect it will trigger many new experiments dealing with heat transport, out-of-equilibrium physics and quantum decoherence.

\subsection*{Acknowledgments}
The authors gratefully acknowledge discussions with M.\ Buttiker, P.\ Degiovanni, C.\ Glattli, P.\ Joyez, A.H.\ MacDonald, F.\ Portier, H.\ Pothier, P.\ Roche, G.\ Vignale. This work was supported by the ANR (ANR-05-NANO-028-03).

\subsection*{Author contributions}
Experimental work and data analysis: C.A., H.l.S. and F.P.; nano-fabrication: C.A. and F.P. with inputs from D.M.; heterojunction growth: A.C. and U.G.; theoretical analysis and manuscript preparation: F.P. with inputs from coauthors; project planning and supervision: F.P.

\subsection*{Additional information}
Supplementary Information accompanies this paper. Correspondence and requests for materials should be addressed to F.P.

\end{document}

\title{Supplementary Information for \\`Non-Equilibrium Edge Channel Spectroscopy in the Integer Quantum Hall Regime'}
\author{C. Altimiras}
\affiliation{CNRS, Laboratoire de Photonique et de Nanostructures
(LPN) - Phynano team, route de Nozay, 91460 Marcoussis, France}
\author{H. le Sueur}
\affiliation{CNRS, Laboratoire de Photonique et de Nanostructures
(LPN) - Phynano team, route de Nozay, 91460 Marcoussis, France}
\author{U. Gennser}
\affiliation{CNRS, Laboratoire de Photonique et de Nanostructures
(LPN) - Phynano team, route de Nozay, 91460 Marcoussis, France}
\author{A. Cavanna}
\affiliation{CNRS, Laboratoire de Photonique et de Nanostructures
(LPN) - Phynano team, route de Nozay, 91460 Marcoussis, France}
\author{D. Mailly}
\affiliation{CNRS, Laboratoire de Photonique et de Nanostructures
(LPN) - Phynano team, route de Nozay, 91460 Marcoussis, France}
\author{F. Pierre}
\affiliation{CNRS, Laboratoire de Photonique et de Nanostructures
(LPN) - Phynano team, route de Nozay, 91460 Marcoussis, France}
\date{October 14, 2009}

\maketitle

\section{Expanded discussion on energy and heat transport by edge excitations}

In this section, after a schematic summary of the experimental principle, we first detail how the energy within the probed electronic excitations can be extracted from the energy distribution function $f(E)$ and we define a generalized temperature $T_{qp}$ for non-equilibrium situations. Second, we detail the link between energy within a system of 1D chiral fermions, such as the `quasiparticles' (i.e. the rigid edge channel displacement excitations), and energy transport. In this case, distribution function and heat current are directly related, without any sample specific parameters. Third, we extract from general power balance considerations the total excess energy current carried by all edge excitations. It is compared to the energy current carried only by the probed quasiparticles. This permits us to rule out any contribution of internal modes to the energy transport within our experimental accuracy.

\begin{figure}[tb]
\includegraphics[width=0.8\columnwidth,clip]{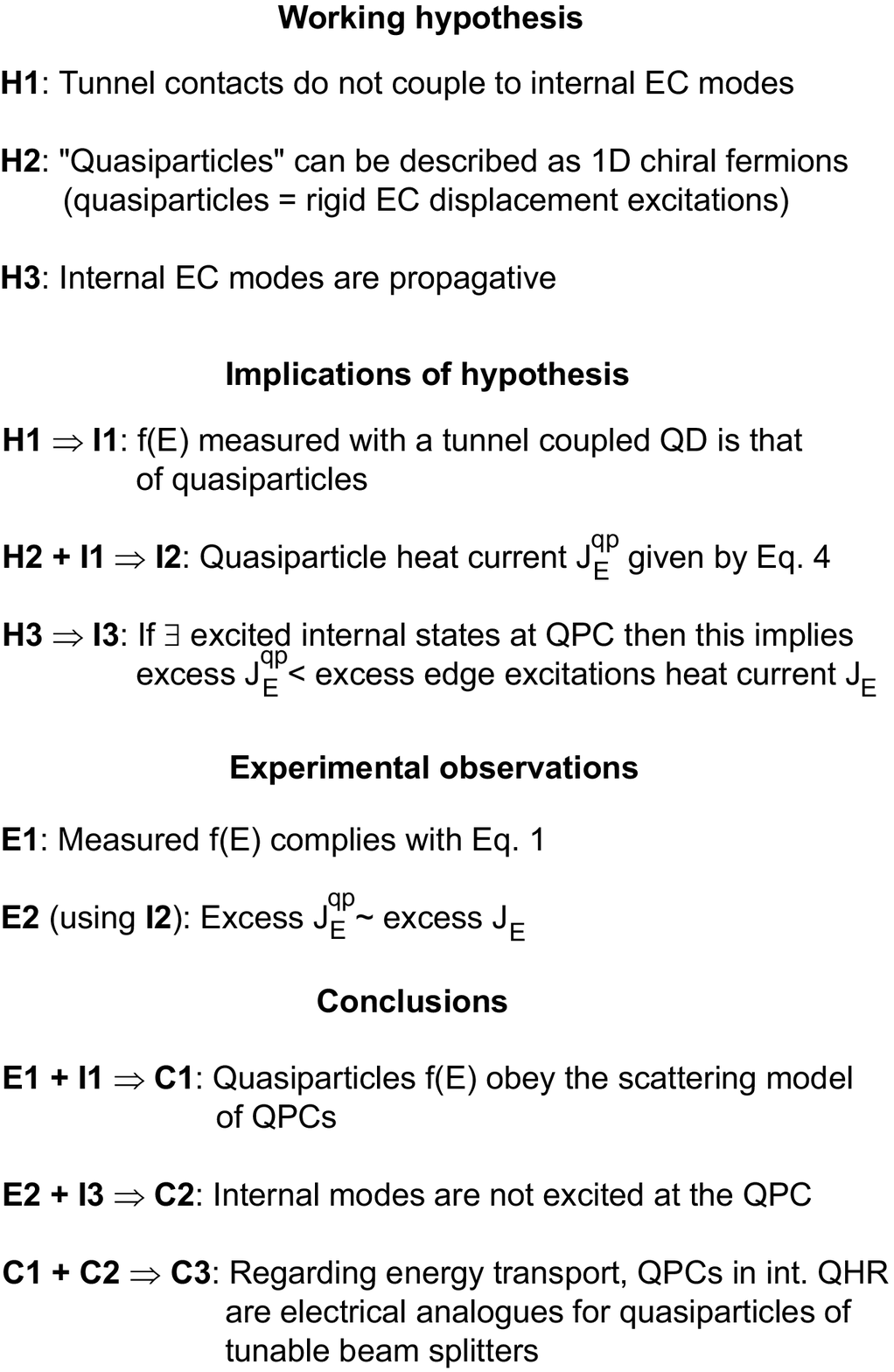}
\caption{{\bfseries Schematic summary of experiment principle.}}
\label{SI_Figsummary}
\end{figure}

\subsection{Schematic summary of experiment principle}

Figure~\ref{SI_Figsummary} recapitulates the logical architecture that permits us to rule out the excitation of internal EC states and to show that QPCs obey the scattering model for quasiparticles \cite{gramespacher1999noneqf}. This demonstrates the quantum optics analogy, regarding energy transport, between a QPC and a tunable beam splitter for quasiparticles.

The hypothesis H1 relies on the linear I-V characteristic for tunnel contacts that is mostly observed (including here) in the integer QHR \cite{aleiner1994nee,conti1996cme,han1997dce,zulicke1999pdt,yang2003ftd}. The hypothesis H2 holds for each edge channel's quasiparticle branch according to \cite{wen1992tesfqh,aleiner1994nee,conti1996cme,han1997dce}. Note that the hydrodynamic approach considers specifically rigid EC displacements \cite{wen1992tesfqh}. The hypothesis H3 is a prediction for internal EC modes \cite{aleiner1994nee,chamon1994er,conti1996cme,han1997dce}.

\subsection{Energy density of non-equilibrium fermions}

At equilibrium and temperature $T$, the energy distribution function $f(E)$ is a Fermi function and the typical energy of electronic excitations is $k_B T$.
In non-equilibrium situations, the distribution function is in general not a Fermi function. Nevertheless, the energy density $E_{qp}$ within a system of fermionic quasiparticles can be obtained from their energy distribution:
\begin{align}
E_{qp}=& \nu \int_{-\infty}^{\mu} (\mu-E)(1-f(E))dE \nonumber
\\& +  \nu \int_\mu ^\infty (E-\mu)f(E)dE,\nonumber \\
=& \nu \int (E-\mu)\delta f(E)dE,\label{Eqp}
\end{align}
with $\nu$ the density of states per unit length and energy, here assumed constant for energies near the local electrochemical potential $\mu$; and $\delta f(E)=f(E)-\theta (\mu-E)$, with $\theta (E)$ the step function, corresponding to variations in the energy distribution relative to the filled Fermi sea. The second line and the right-hand side of the first line correspond to the energy density contribution of, respectively, electron and hole like quasiparticles. In the important case of a Fermi function at temperature $T$, the energy density obtained from Supplementary equation~\ref{Eqp} is the standard textbook value \cite{pinesnozieres1966tql}:
\begin{equation}
E_{qp}=\frac{\pi^2}{6}\nu (k_B T)^2. \label{EdeT}
\end{equation}
In practice, it is more convenient to use the electronic temperature $T_{qp}$ generalized to non-equilibrium situations, which is independent of the density of states, rather than the energy density $E_{qp}$:
\begin{equation}
T_{qp} \equiv \sqrt{\frac{6 (E_{qp}/\nu)}{\pi^2}}/k_B. \label{Tqp}
\end{equation}
The energy density and generalized electronic temperature can be computed analytically for the `double step' energy distributions detailed by Eq.~(2). One finds:
\begin{align}
E_{qp}/\nu =& \frac{(\pi k_B T)^2}{6}+\tau (1-\tau) \frac{(e \delta V_D)^2}{2},\label{E2step} \\
T_{qp}=& \sqrt{T^2 + 3 \tau (1-\tau) \left( \frac{e \delta V_D}{\pi k_B}\right)^2 } .\label{T2step}
\end{align}

\subsection{Energy transport by 1D chiral fermions}

We expect from the rigid displacement model \cite{zulicke1999pdt,conti1996cme,han1997dce,aleiner1994nee} supported by tunneling density of states experiments, that the energy distribution, measured here with a weakly coupled quantum dot, is that of rigid displacement excitations (the quasiparticles) and not of internal edge channel excitations. Moreover, it was shown that the quasiparticles of each edge channel can be mapped onto a branch of 1D chiral fermions \cite{wen1992tesfqh,aleiner1994nee,conti1996cme,han1997dce}. Therefore, the measured energy current $J_E^{qp}$ along an edge channel corresponds to that of quasiparticles and reads:
\begin{equation}
J_E^{qp}=v E_{qp}=\frac{\pi^2}{6 h} (k_B T_{qp})^2, \label{JE}
\end{equation}
with $v$ the drift velocity and $h$ the Plank constant. Note first that the right-hand side expression of heat current holds even if the density of states $\nu$ depends on energy, as long as the very robust 1D velocity-density of states cancelation $v \nu=1/h$ is obeyed \cite{sivan1986mlfftt}. Second, as expected the above expression obeys the Wiedemann-Franz law. Third, we point out that the energy flow is directly given by the generalized temperature $T_{qp}$, without any sample specific parameters. Therefore, by measuring the energy distribution function in our experiment, we probe quantitatively the quasiparticle energy current.

Assuming the quasiparticle distribution function is given by Equation~(2), the corresponding energy flow is:
\begin{equation}
J_E^{qp}=\frac{\pi^2}{6 h}(k_B T)^2 + \tau (1-\tau) \frac{(e \delta V_D)^2}{2 h}. \label{JEdata}
\end{equation}

\subsection{Excess energy transport by all edge excitations and comparison with that by quasiparticles}

\begin{figure}[tbh]
\includegraphics[width=1\columnwidth,clip]{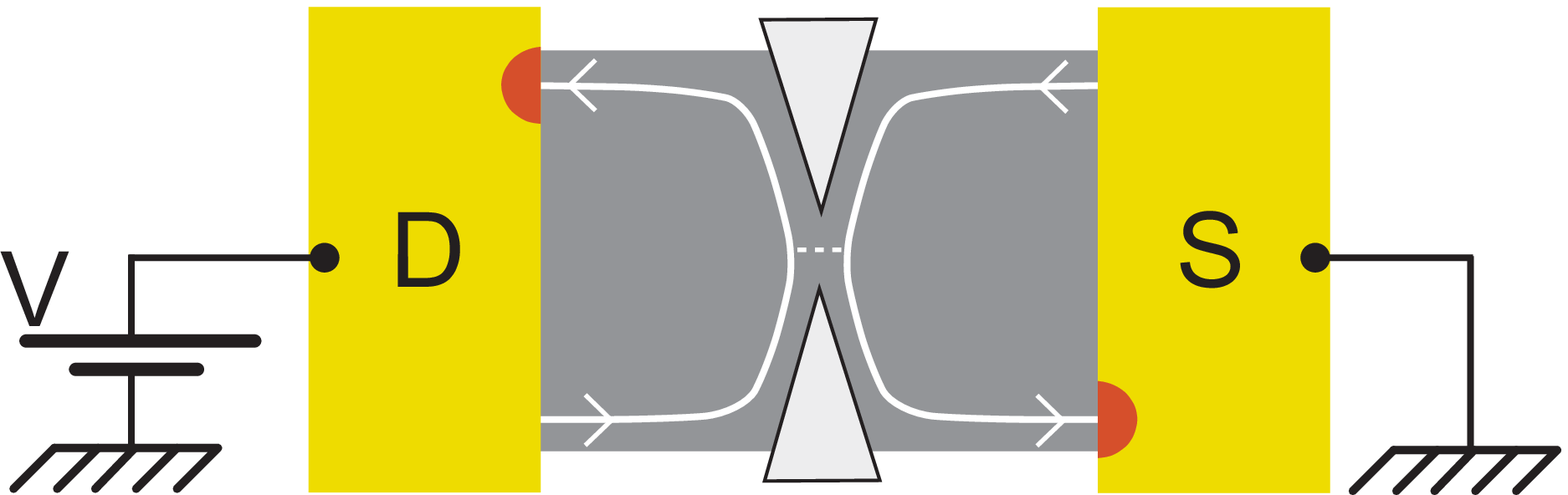}
\caption{{\bfseries Simplified schematic circuit used for power balance considerations.} The Landau level filling factor is here set to one. The edge channel is shown as a white line and the propagation direction of electronic excitations is indicated by an arrow. The dissipated power $V^2G_{QPC}=V^2\tau e^2/h$ is injected into the source (S) and drain (D) electrodes by the corresponding incoming edge channels.} \label{SI_schematic}
\end{figure}

We consider the simplified circuit at filling factor 1 (one edge channel) shown in Supplementary figure~\ref{SI_schematic}.

Here, the total power $P=V^2 \tau e^2/h$, with $\tau$ the quantum point contact transmission, is dissipated in the source (S) and drain (D) electrodes (nearby the hot spots shown as red areas in Supplementary figure~\ref{SI_schematic}, see e.g. \cite{klass1991idq}). It can be decomposed into two contributions:
\begin{equation}
P=V^2 \tau e^2/h=P_{\delta\mu}+P_{edge}. \label{P}
\end{equation}
The first one ($P_{\delta\mu}$) corresponds to the energy injected into the drain and source electrodes due to the electrochemical potential difference $\delta\mu$ between electrode and corresponding incoming edge. The edge electrochemical potential is defined as that of a floating electrode inserted in its path, in the spirit of the `measurement reservoir' model (see e.g. \cite{sivan1986mlfftt}). At unity transmission $\tau=1$, this `electrochemical power' is the only contribution to the dissipated power $P_{\delta\mu} (\tau=1)=P=(eV)^2/h$. In general, the electrochemical power injected by each edge in its output electrode is $(\delta\mu)^2/2h$. At arbitrary transmission $\tau$, the electrochemical potential difference at the input of both the source and drain electrodes is $|\delta\mu|=\tau e |V|$ and one finds:
\begin{equation}
P_{\delta\mu}=(\tau e V)^2/h. \label{Pmu}
\end{equation}
The second contribution $P_{edge}=2(J_E^{in}-J_E^{out})$ corresponds to the difference between the incoming $J_E^{in}$ and outgoing $J_E^{out}$ energy current carried on by \emph{all edge excitations}, respectively in and out the corresponding electrode. The factor two here accounts for the two electrodes. Note that $P_{edge}$ corresponds to the amount of energy that would be absorbed by two floating reservoirs inserted along the path of the edges incoming to source and drain electrodes and thermalized at the same temperature as their corresponding electrode. This contribution vanishes at zero transmission and also at unity transmission, as long as the drain and source electrodes are at the same temperature since in that case $J_E^{in}=J_E^{out}$. At intermediate transmissions, $P_{edge}$ is obtained from Supplementary equations~\ref{P} and \ref{Pmu}:
\begin{equation}
P_{edge}=P-P_{\delta\mu}=\tau (1-\tau) \frac{(e \delta V_D)^2}{h}. \label{Pedge}
\end{equation}
This last quantity is identical to the one obtained from the double step distribution function (using $J_E^{in}$ given by Supplementary equation~\ref{JEdata} and $J_E^{out}=\frac{\pi^2}{6 h}(k_B T)^2$). Therefore the observed agreement between measured quasiparticle $f(E)$ and the prediction of Eq.~2 already implies that the excess energy current is carried on by quasiparticles and not internal edge channel modes. However, the most straightforward evidence is to extract the energy current directly from the measured quasiparticle energy distribution function using Supplementary equations~\ref{Eqp}, \ref{Tqp} and \ref{JE}. The internal modes being propagative, the observation that excess quasiparticle energy current and full excess edge current are similar (see Figures~3d and 4d) implies internal edge channel states are not excited by a voltage biased QPC of arbitrary transmission within our experimental accuracy.

\section{Supplementary Data}

\begin{figure}[tb]
\includegraphics[width=0.85\columnwidth,clip]{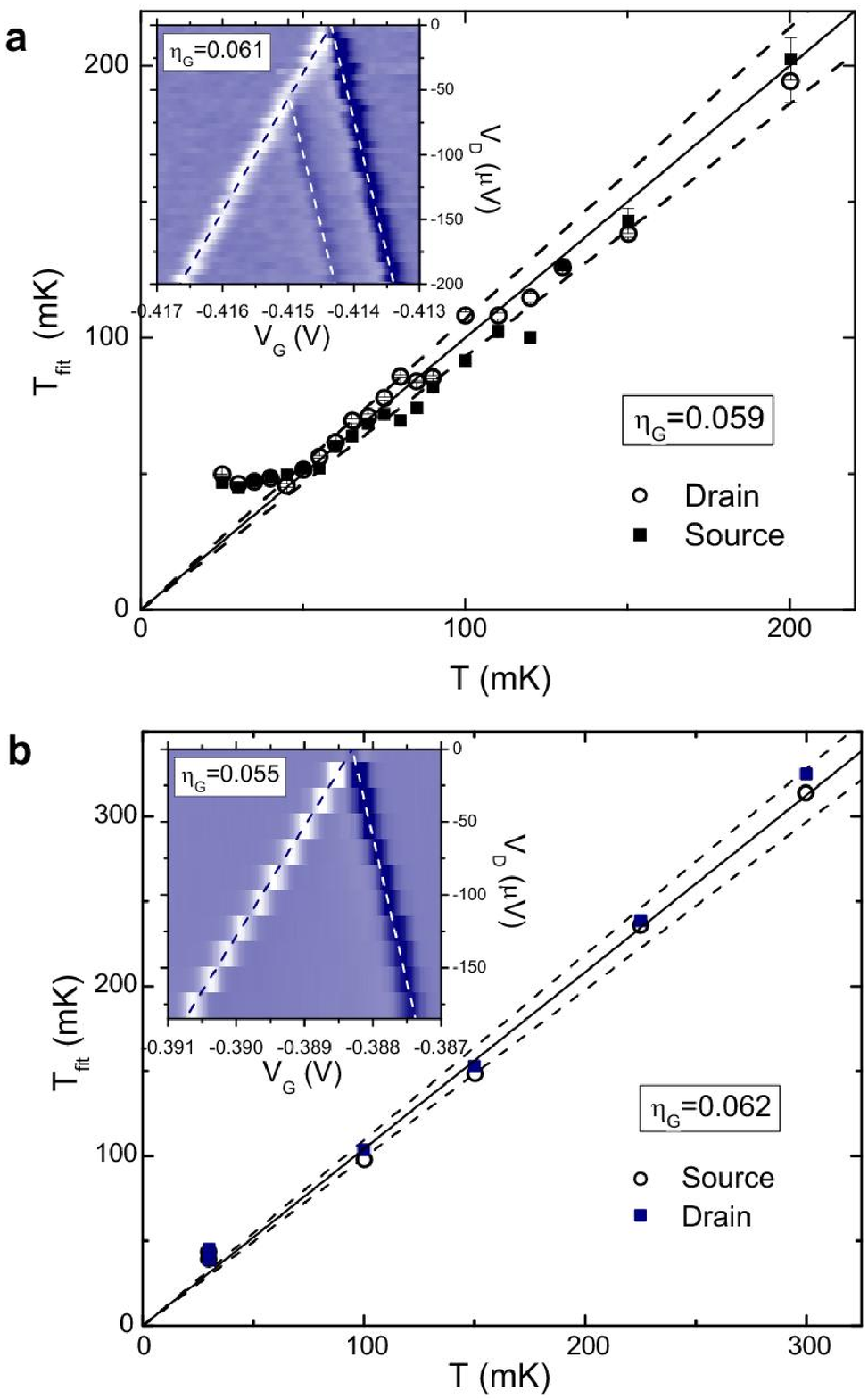}
\caption{{\bfseries Quantum dot calibration in cooldowns two and three.} This figure complements Figure~2 that focused on cooldown one. ({\bfseries a}) and ({\bfseries b}) are calibration data for cooldowns two and three, respectively. Errors in $T_{fit}$ within $\pm7\%$ for cooldown two ({\bfseries a}) and within $\pm5\%$ for cooldown three ({\bfseries b}) are enclosed between the black dashed lines. Note that an additional QD level of relative energy $-56~\mu e$V for cooldown two is visible in the top left inset of ({\bfseries a}).} \label{SI_datacalib}
\end{figure}

\begin{figure}[tbh]
\includegraphics[width=1\columnwidth,clip]{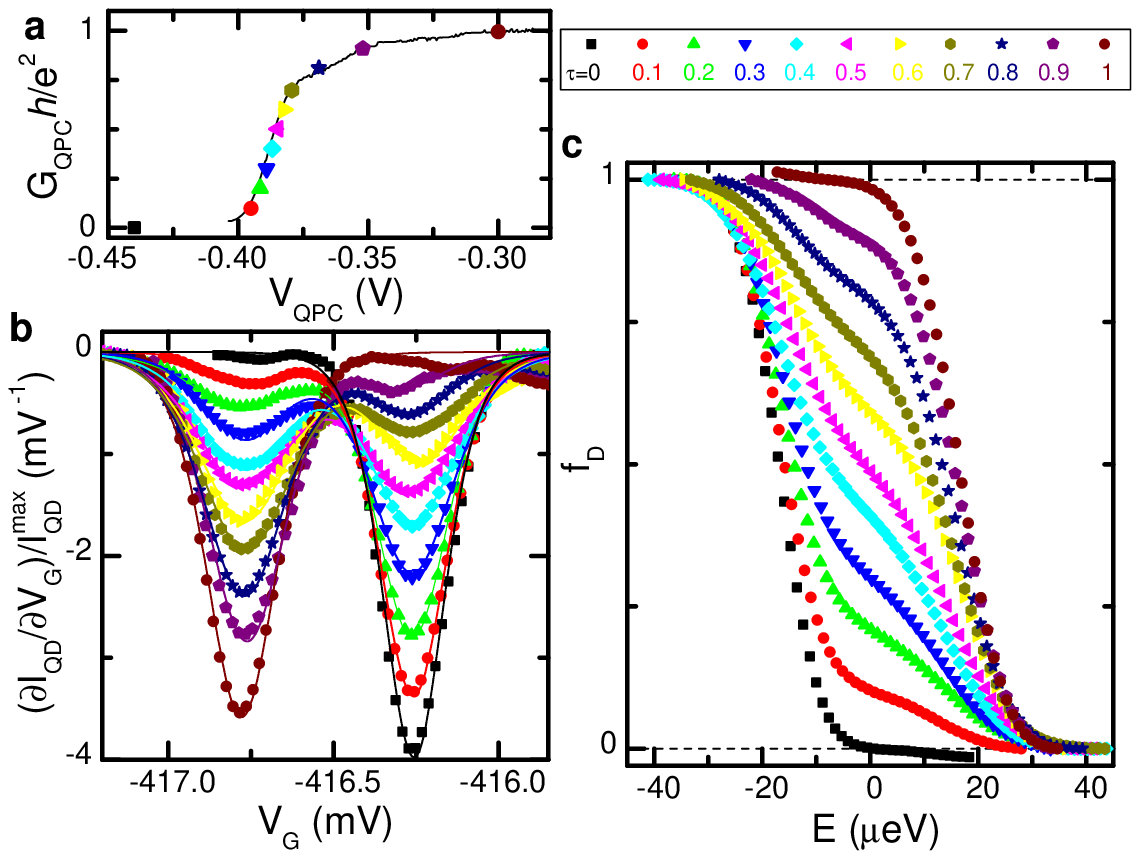}
\caption{{\bfseries Spectroscopy of an edge channel tuned out-of-equilibrium with the conductance of a QPC in cooldown two.} The QPC's voltage bias is here fixed to $\delta V_D \equiv V_{D1}-V_{D2}=-36~\mu$V. {\bfseries a}, Measured $G_{\mathrm{QPC}}$ versus $V_{\mathrm{QPC}}$ applied to the lower left metal gate in Figure~1c. Symbols are data points obtained during the corresponding EC spectroscopy. The continuous line was measured with $\delta V_D=0$. {\bfseries b}, Measured $\partial I_{\mathrm{QD}}/\partial V_G$ (symbols) for the QD-drain contribution (data have been aligned in $V_G$ and normalized to $I_{\mathrm{QD}}^{max}$). Continuous lines are fits assuming $f_D(E)$ is the weighted sum of two Fermi functions. The detailed set of used fit parameters is given in Supplementary table~\ref{SItabC2}. {\bfseries c}, Energy distributions obtained by integrating the data in ({\bfseries b}) and using $\eta_G=0.059$.}
\label{SIfigC2}
\end{figure}

\begin{figure}[tbh]
\includegraphics[width=1\columnwidth,clip]{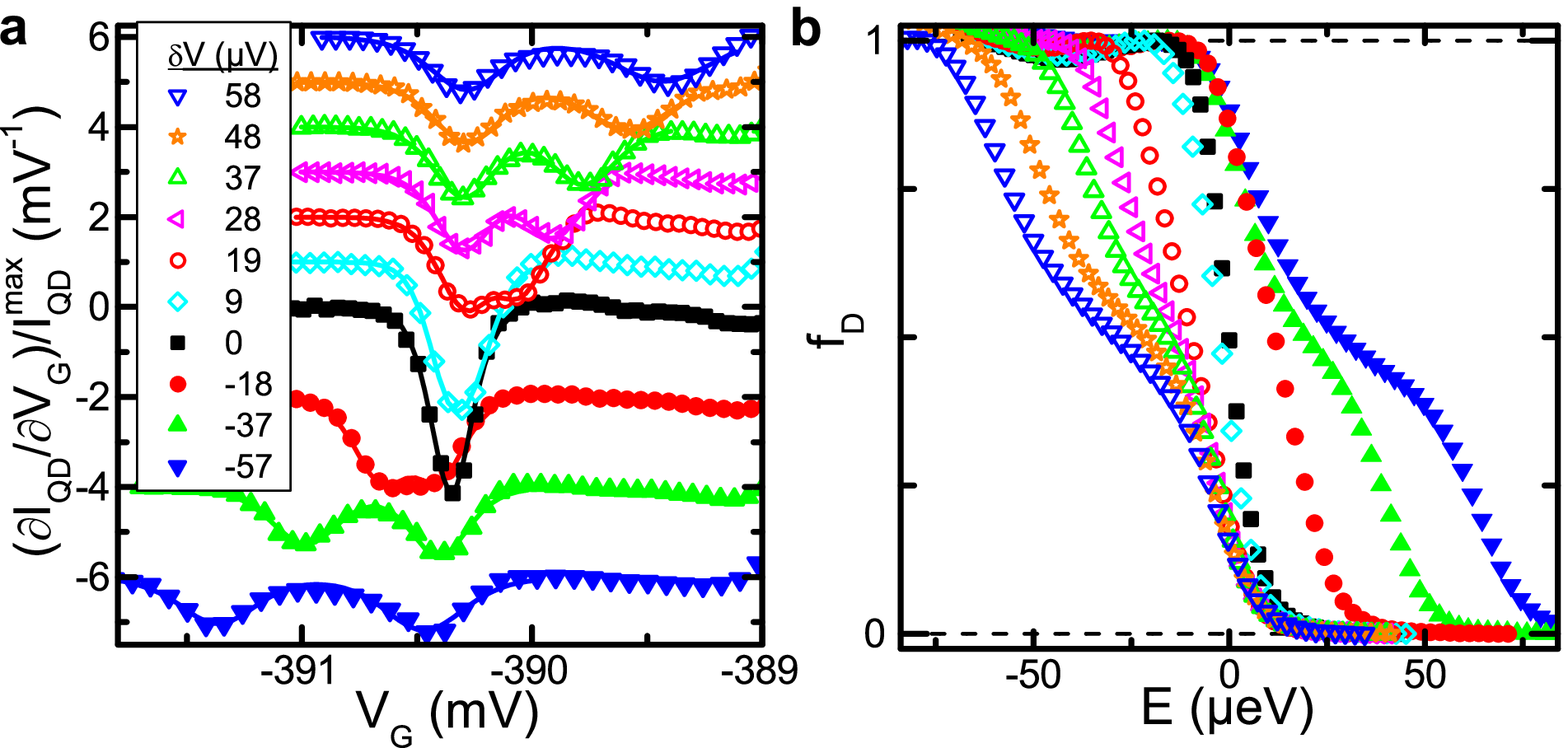}
\caption{{\bfseries Spectroscopy of an edge channel tuned out-of-equilibrium with the voltage across a QPC in cooldown three.}
The QPC's conductance is here fixed to $G_{\mathrm{QPC}}=0.5 e^2/h$. {\bfseries a}, Measured $\partial I_{\mathrm{QD}}/\partial V_{G}$ (symbols) for the QD-drain contribution. Data have been shifted vertically for clarity, and horizontally to align the peak corresponding to the fixed potential $V_{D2}$. Continuous lines are fits assuming $f_D(E)$ is the weighted sum of two Fermi functions (see Supplementary table~\ref{SItabC3}). {\bfseries b}, Energy distributions obtained by integrating the data in ({\bfseries a}) and using $\eta_G=0.062$.}
\label{SIfigC3}
\end{figure}

In this section we present supplementary data regarding cooldowns two and three, that were not shown in the article.

The quantum dot calibration data are shown in Supplementary figure~\ref{SI_datacalib}.

Data obtained in cooldown two with a fixed bias voltage $\delta V_D=-36~\mu$V and several values of the QPC conductance are shown in Supplementary figure~\ref{SIfigC2}. These data correspond to symbols ($+$) in Figure~3c,d. Note that here $V_{D2}$ and the inner edge channel potential are set to $-129~\mu$V.

Data obtained in cooldown three with a fixed QPC conductance $G_{QPC}\simeq0.5e^2/h$ and several values of the QPC voltage bias $\delta V_D$ are shown in Supplementary figure~\ref{SIfigC3}. As for cooldown one, $V_{D2}$ and the inner edge channel potential are here fixed to $-88~\mu$V.

\section{Supplementary information on methods}

\subsection{Measured sample and experimental techniques}

The sample was realized in a standard GaAs/Ga(Al)As two dimensional electron gas 105~nm below the surface, of density $2~10^{15}~\mathrm{m}^{-2}$, Fermi energy $80~$K and mobility $250~\mathrm{m}^2V^{-1}s^{-1}$. Note that the GaAs/Ga(Al)As heterojunction used here is the same one as that used formerly to perform the Mach-Zehnder experiments with edge states reported in \cite{roulleau2008lphi,roulleau2008noisedephasing}. The silicon (dopant) concentration in the heterojunction was adjusted to optimize the Hall resistance quantization.

Conductance measurements were performed in a dilution refrigerator of base temperature 30~mK. All measurement lines were filtered by commercial $\pi$-filters at the top of the cryostat. At low temperature, the lines were carefully filtered and thermalized by arranging them as 1~m long resistive twisted pairs ($300~\Omega /$m) inserted inside 260~$\mu$m inner diameter CuNi tubes tightly wrapped around a copper plate screwed to the mixing chamber. The sample was further protected from spurious high energy photons by two shields, both at base temperature.

The sample was current biased by a voltage source in series with a $10~$M$\Omega$ or $100~$M$\Omega$ polarization resistance at room temperature. Taking advantage of the well defined quantum Hall resistance (12.906~k$\Omega$), currents across the sample were converted on-chip into voltages and measured with low noise room temperature voltage amplifiers. To limit artifacts by slowly moving charges nearby the QD, we systematically measured several successive gate voltage sweeps $I_{\mathrm{QD}}(V_G)$, checked that the data fall on top of each other, and verified that the sum rule $\int (\partial I_{\mathrm{QD}}/\partial V_G) dV_G \simeq 0$ is obeyed.

\subsection{Quantum dot theoretical model}

The QD modeling follows the master equation approach for incoherent sequential tunneling. The derivation detailed in \cite{averin1991tsec} stays valid for non-Fermi distribution functions. To obtain the simple expression of Equation~1 we assumed only one QD level contributes to $I_{\mathrm{QD}}$, and neglected the energy dependence of the electrodes tunneling density of states and of the tunnel rates in and out the QD. In practice, the validity of these hypotheses are checked with the non-linear QD characterization and by comparing mixing chamber temperatures with fit temperatures obtained within this framework (see Figure~2 and Supplementary figure~\ref{SI_datacalib}). Note that, beyond sequential tunneling, a similar relationship between $I_{\mathrm{QD}}$ and $f_{S,D}$ holds in the resonant transport regime but with distribution functions artificially smoothed by the energy width of the QD level.

\subsection{Experimental procedures}
\subsubsection{Extraction of the energy distribution function}
The distribution functions $f(E)$ are obtained by integrating with $V_G$ the normalized data $-(\partial I_{\mathrm{QD}}/\partial V_G)/I_{\mathrm{QD}}^{max}$ from a charge stability zone ($I_{\mathrm{QD}}=0$) on the negative side up to $V_G=-(E-E_0)/e\eta_G$.

\subsubsection{Extraction of the electrochemical potential}
The electrochemical potential $\mu$ enters in the evaluation of the energy density (Supplementary equation~\ref{Eqp}), and consequently in the heat current $J_E^{qp}$ and in the generalized quasiparticle temperature $T_{qp}$. The parameter $\mu$ is here obtained very directly from the measured $f(E)$ using
\begin{equation}
\mu=E_{min}+\int_{E_{min}}^{E_{max}}f(E)dE, \label{mu}
\end{equation}
with $E_{min}$ ($E_{max}$) an energy under (above) which we assume $f(E)=1$ ($f(E)=0$). Note that $\mu$ is obtained up to the same unknown constant offset $E_0$ as $f(E)$, however this has no impact since only the difference $E-\mu$ plays a role.

\subsubsection{Estimation of uncertainties on the lever arm $\eta_G$}
Uncertainties in the gate voltage-to-energy lever arm conversion factor $\eta_G$ are mostly responsible for our error bars on energy related quantities (e.g. Figs.~3d and 4c,d).

This parameter is extracted from two independent calibration procedures:
First, we perform a non-linear characterization of the QD \cite{kouwenhoven1997etqd} and extract $\eta_G$ from the slopes of the Coulomb diamond $\partial I_{\mathrm{QD}}/\partial V_G(V_G,V_\mathrm{D})$ (hereafter called procedure~1).
Second, we measure $\partial I_{\mathrm{QD}}/\partial V_G(V_G)$ at several temperatures and extract $\eta_G$ from the scaling between fit temperatures (using Fermi functions in Equation~1) and measured mixing chamber temperatures (hereafter called procedure~2).

In procedure~1, uncertainties on $\eta_G$ are obtained from the change in slopes corresponding to displacements equal to the full width at half maximum of the peaks at the highest drain-source voltage. This gives $\eta_G=0.052\pm 9\%$, $0.062\pm 8\%$ and $0.055\pm 9\%$ for, respectively, cooldowns one, two and three. Note that these uncertainties correspond to plus and minus the full width at half maximum corresponding to the effective electronic temperature 50~mK (i.e. $0.05\times3.5k_B/e\simeq15~\mu$V) divided by the maximum drain voltage $V_D$ applied.

In procedure~2, uncertainties in $\eta_G$ are obtained from the dispersion in fit temperatures $T_{fit}$ around measured temperatures $T$ of the dilution refrigerator mixing chamber. We evaluate roughly the uncertainty by finding the range of $\eta_G$ that permits us to account for most $T_{fit}$ at $T>50~$mK. The reader can get a direct idea of the used uncertainties in Figure~2 and Supplementary figure~\ref{SI_datacalib}, where the expected fit temperatures using the considered extremal values of $\eta_G$ are shown as dashed lines. This gives $\eta_G=0.057\pm 10\%$, $0.059\pm 7\%$ and $0.062\pm 5\%$ for, respectively, cooldowns one, two and three.

In the article, we have chosen to use the values of $\eta_G$ and the associated uncertainties extracted using procedure~2, in the same experimental configuration as to measure unknown $f(E)$s.

\subsubsection{Estimation of error bars}
The error bars in the figures take into account the finite signal to noise and reproducibility (i.e. the standard error in the average value), and, if it applies, uncertainties in the lever arm.

Regarding the finite signal to noise and reproducibility our approach is very straightforward. We perform a statistical analysis on the considered quantity extracted from up to a hundred different $V_G$ sweeps in the exact same experimental configuration. The corresponding standard error is then plus/minus the mean deviation per sweep divided by the square root of the number of sweeps. This is the only contribution for the parameter $\tau_{fit}$ whose typical error bars are found to be about 0.02, small compared to the symbol size in Fig.~3c. Note that in practice, we acquired the large number of sweeps necessary for an accurate statistical analysis in only a few realizations per experiment. In other realizations of a given experiment (i.e. when changing only the QPC conductance or the applied voltage bias), we assumed that the observed mean deviation per sweep is unchanged and estimate the standard error using the corresponding number of sweeps (generally more than five).

It turns out that, except for the immune $\tau_{fit}$, error bars are mostly dominated by uncertainties on the lever arm $\eta_G$ and therefore are proportional to the overall energy (namely, $T_{qp}$ or $\delta V_D$). Indeed, the standard error on $T_{qp}$ was always found smaller than 1~mK for cooldowns one and three, and 4~mK for cooldown two. Nevertheless, the full error bars shown in figures include both contributions, taken as independent from each other.

\begin{table}
\begin{tabular}{|c|c|c|c|c|c|c|c|c|}
  \hline
  $G_{QPC}$ $(e^2/h)$ & $\tau_{fit}$ & $T_{D1} (K)$ & $T_{D2} (K)$  \\
  \hline
  0 & 0 & 0.042 & 0.042 \\
  \hline
  0.15 & 0.18 & 0.158 & 0.042 \\
  \hline
  0.26 & 0.25 & 0.078 & 0.046 \\
  \hline
  0.35 & 0.33 & 0.071 & 0.049 \\
  \hline
  0.43 & 0.42 & 0.067 & 0.050 \\
  \hline
  0.51 & 0.48 & 0.062 & 0.054 \\
  \hline
  0.59 & 0.58 & 0.064 & 0.054 \\
  \hline
  0.59 & 0.61 & 0.067 & 0.056 \\
  \hline
  0.65 & 0.65 & 0.062 & 0.056 \\
  \hline
  0.72 & 0.73 & 0.061 & 0.053 \\
  \hline
  0.78 & 0.77 & 0.056 & 0.059 \\
  \hline
  0.88 & 0.86 & 0.053 & 0.073 \\
  \hline
  1 & 1 & 0.045 & 0.045 \\
  \hline
\end{tabular}
\caption{{\bfseries Summary of parameters used to fit the data shown in Figure~3b (cooldown one).} The conductance $G_{QPC}$ is measured. Note that near zero and full transmission, the fit temperature of the small corresponding peak is not very significant. In order to focus on the fit parameter $\tau_{fit}$, we chose to fix $\eta_G \delta V_{G}=30~\mu$V. If $\eta_G \delta V_{G}$ is set free, we find values within $30\pm1~\mu$V, except at $G_{QPC}=0.9e^2/h$ where $\eta_G \delta V_{G}=39~\mu$V. Note that $\tau_{fit}$ is not affected more than $\pm0.03$ by whether $\eta_G \delta V_{G}$ is fixed or free.}
\label{SItabC1tau}
\end{table}

\begin{table}
\begin{tabular}{|c|c|c|c|c|c|}
  \hline
  $\delta V_{D}$ $(\mu V)$ & $T_{D1}$ (K) & $T_{D2}(K)$ & $\eta_G \delta V_{G}$ ($\mu V$) \\
  \hline
  -18 & 0.049 & 0.049 & -12 \\
  \hline
  0 & 0.040 & 0.040 & 0 \\
  \hline
  18 & 0.048 & 0.045 & 13 \\
  \hline
  27 & 0.054 & 0.054 & 20  \\
  \hline
  36 & 0.061 & 0.056 & 29 \\
  \hline
  45 & 0.074 & 0.063 & 38 \\
  \hline
  54 & 0.076 & 0.073 & 46 \\
  \hline
\end{tabular}
\caption{{\bfseries Summary of parameters used to fit the data shown in Figure~4b (cooldown one).} Here $\delta V_D$ is the applied voltage bias and the parameter $\tau_{fit}$ is set to the measured $G_{QPC}h/e^2=0.58$.}
\label{SItabC1dV}
\end{table}

\begin{table}
\begin{tabular}{|c|c|c|c|c|c|}
  \hline
  $G_{QPC}$ $(e^2/h)$ & $\tau_{fit}$ & $T_{D1}=T_{D2}$(K) & $\eta_G \delta V_{G}$ ($\mu V$)  \\
  \hline
  0 & 0 & 0.043 & -36 \\
  \hline
  0.10 & 0.08 & 0.046 & -30 \\
  \hline
  0.20 & 0.18& 0.050 & -31 \\
  \hline
  0.30 & 0.28 & 0.057 & -30  \\
  \hline
  0.40 & 0.41 & 0.062 & -30 \\
  \hline
  0.50 & 0.49 & 0.065 & -30 \\
  \hline
  0.60 & 0.61 & 0.063 & -31 \\
  \hline
  0.70 & 0.71 & 0.063 & -30 \\
  \hline
  0.81 & 0.80 & 0.057 & -30 \\
  \hline
  0.91 & 0.89 & 0.054 & -28 \\
  \hline
  1 & 1 & 0.049 & -36 \\
  \hline
\end{tabular}
\caption{{\bfseries Summary of parameters used to fit the data shown in Supplementary figure~\ref{SIfigC2}b (cooldown 2).} The conductance $G_{QPC}$ is measured. The applied QPC voltage bias is here $\delta V_D=-36~\mu$V. The lever arm is $\eta_G=0.059$.}
\label{SItabC2}
\end{table}

\begin{table}
\begin{tabular}{|c|c|c|c|c|c|}
  \hline
  $\delta V_{D} (\mu V)$ & $T_{D1}$ (K) & $T_{D2}$(K) & $\eta_G \delta V_{G}$ ($\mu V$)  \\
  \hline
  -57 & 0.089 & 0.077 & -56 \\
 \hline
  -37 & 0.073 & 0.061 & -36 \\
  \hline
  -18 & 0.055 & 0.054 & -15 \\
  \hline
  0 & 0.044 & 0.044 & 0 \\
  \hline
  9 & 0.046 & 0.043 & 7 \\
  \hline
  19 & 0.057 & 0.052 & 15 \\
  \hline
  28 & 0.064 & 0.055 & 24  \\
  \hline
  37 & 0.073 & 0.060 & 32 \\
  \hline
  48 & 0.088 & 0.070 & 44 \\
  \hline
  58 & 0.094 & 0.080 & 53 \\
  \hline
\end{tabular}
\caption{{\bfseries Summary of parameters used to fit the data shown in Supplementary figure~\ref{SIfigC3}a (cooldown 3).} Here $\delta V_D$ is the applied voltage bias and the parameter $\tau_{fit}$ is set to the measured $G_{QPC}h/e^2=0.5$. The lever arm is $\eta_G=0.062$.}
\label{SItabC3}
\end{table}

\subsubsection{Fit procedures}

We fitted the measured $(\partial I_{\mathrm{QD}}/\partial V_G)/I_{QD}^{max}$ using Equations~1 and 2, with Fermi functions for $f_{D1}$ and $f_{D2}$. The full set of fit parameters is not only constituted of $\tau_{fit}$ and $\eta_G \delta V_G$ shown in Figures~3c and 4c, respectively. It also includes the temperatures $T_{D1}$ and $T_{D2}$ of the corresponding Fermi functions. We recapitulate the full set of used fit parameters in Supplementary tables~\ref{SItabC1tau}, \ref{SItabC1dV}, \ref{SItabC2} and \ref{SItabC3}.

\begin{figure}[bt]
\includegraphics[width=0.7\columnwidth,clip]{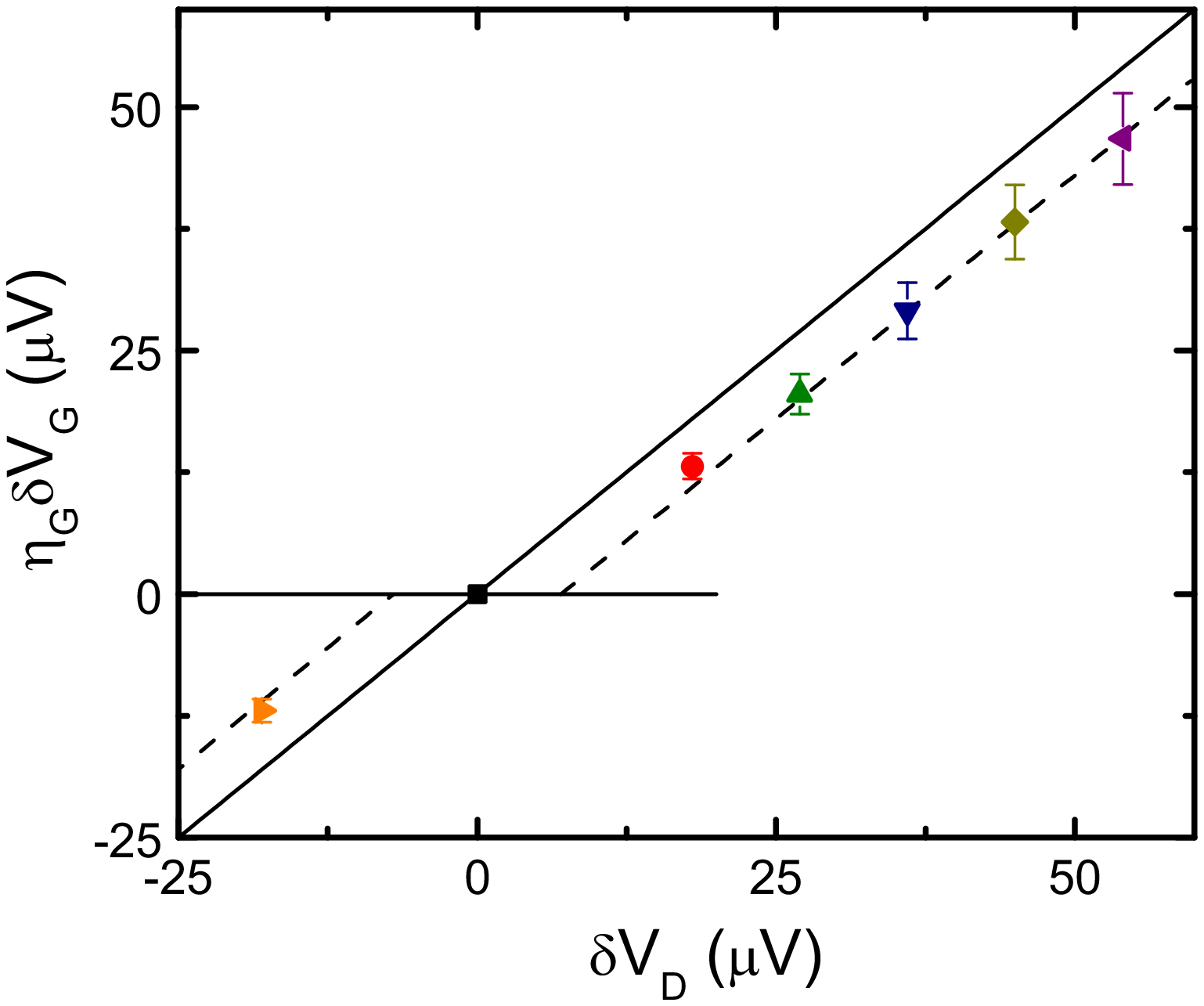}
\caption{{\bfseries Possible explanation with additional QD levels of observed small deviations on cooldown one.} Data shown as symbols are those obtained in cooldown one and also displayed in Figure~4c. Dashed lines correspond to the slope predicted by $\eta_G=0.057$ but shifted horizontally by $\pm 7~\mu$V. Such shifts are compatible with the presence of a second active QD level located at $\varepsilon \approx -10~\mu e$V, below the first level.} \label{SI_addlevel}
\end{figure}

\section{Supplementary discussion regarding deviations between data and free 1D chiral fermions theory}

In this section, we discuss two possible explanations of the observed deviations between the data and the free 1D chiral fermions predictions that can be seen in Figures~3d, 4c and 4d. The first one relies on additional active quantum dot levels. The second one on energy redistribution with the inner edge channel.

\subsection{Effect of additional quantum dot levels}

In the presence of more than one active quantum dot level, the straightforward Equation~1 does not hold. In the sequential tunneling regime, the master equation approach applies \cite{kouwenhoven1997etqd}. This can result in non-intuitive phenomena. In particular, it was shown that the gate voltage position of a Coulomb peak can be shifted by about its width when the temperature changes \cite{bonet2002sre}. We found a similar phenomenon could change the gate voltage separation between the two dips observed in presence of a double step energy distribution.

By solving the master equation with non-equilibrium energy distributions, using a slightly modified version of the code provided by \cite{bonet2002sre}, and with a second level of energy $\varepsilon$ nearby the first active level, we could reproduce the observed difference between $\eta_G \delta V_G$ and $\delta V_D$ shown in Figure~4d and Supplementary figure~\ref{SI_addlevel}. Such second level for cooldown one is not visible in the top left inset of Figure~2 but its existence is suggested by data taken in slightly different conditions (not shown) and by the observed $T_S>T_D$ at $T=30~$mK (see Figure~2). The presence of such a level can reduce $|\delta V_{G}|$ by a constant offset for $|e \delta V_{D}|>|\varepsilon|$ if asymmetrically coupled to the source and drain electrodes. Supplementary figure~\ref{SI_addlevel} shows $\eta_G \delta V_G$ extracted from cooldown one (data also shown in Figure~4c) together with dashed lines of slopes as expected from $\eta_G=0.057$ but offset by $\pm7~\mu$V which is compatible with a second level located at $\varepsilon \approx -10~\mu e$V, below the first level.

\subsection{Effect of energy redistribution with inner edge channel}

More recent data that we obtained using the same experimental configuration show that an energy redistribution exists between the outer and inner edge channels on length scales larger than $2~\mu$m \cite{lesueur2009relaxqhr}. This effect, although probably small for the short $0.8~\mu$m propagation length considered here, could explain the also small deviation observed between the free 1D chiral fermions model and our data in Figures~3d and 4d. Another indication that energy relaxation along the edge is not fully negligible is the observed increase in the fit temperatures $T_{D1}$ and $T_{D2}$ with injected power (see Supplementary tables~\ref{SItabC1tau}, \ref{SItabC1dV}, \ref{SItabC2} and \ref{SItabC3}).

%\bibliographystyle{nature}
%\bibliography{biblio}